%% file: main.tex
\documentclass[sigconf]{acmart}

\usepackage{booktabs} 
\usepackage[normalem]{ulem}
\usepackage{wrapfig}
\usepackage{xcolor}
\usepackage{array}
\usepackage{multirow}
\usepackage[inline]{enumitem}
\usepackage{balance}

\newcommand{\CQA}[1]{#1}
\newcommand{\SE}[1]{\textsc{#1}}

\copyrightyear{2018}
\acmYear{2018} 
\setcopyright{iw3c2w3}
\acmConference[WWW 2018]{The 2018 Web Conference}{April 23--27, 2018}{Lyon, France}
\acmBooktitle{WWW 2018: The 2018 Web Conference, April 23--27, 2018, Lyon, France}
\acmPrice{}
\acmDOI{10.1145/3178876.3186037}
\acmISBN{978-1-4503-5639-8/18/04}

\begin{document}
\title[The Size Conundrum]{The Size Conundrum: Why Online Knowledge Markets Can Fail at Scale}

\author{Himel Dev}
\affiliation{\institution{University of Illinois (UIUC)}}
\email{hdev3@illinois.edu}

\author{Chase Geigle}
\affiliation{\institution{University of Illinois (UIUC)}}
\email{geigle1@illinois.edu}

\author{Qingtao Hu}
\affiliation{\institution{University of Illinois (UIUC)}}
\email{qhu13@illinois.edu}

\author{Jiahui Zheng}
\affiliation{\institution{Peking University}}
\email{alicehz@pku.edu.cn}

\author{Hari Sundaram}
\affiliation{\institution{University of Illinois (UIUC)}}
\email{hs1@illinois.edu}

\renewcommand{\shortauthors}{H. Dev et al.}

\input{0_abstract}

%
%



\maketitle
\input{1_introduction}

\input{2_related}

\input{3_problem}
\input{4_model}

\input{5_dataset}
\input{6_empirical}

\input{7_interpretation}

\input{8_diseconomies}

\input{9_discussion}

\input{x_conclusion}

\balance

\bibliographystyle{ACM-Reference-Format}
\bibliography{www_2018}

\end{document}

%% file: 0_abstract.tex
\begin{abstract}
In this paper, we interpret the community question answering websites on the \CQA{StackExchange} platform as knowledge markets, and analyze how and why these markets can fail at scale. A knowledge market framing allows site operators to reason about market failures, and to design policies to prevent them. Our goal is to provide insights on large-scale knowledge market failures through an interpretable model. We explore a set of interpretable economic production models on a large empirical dataset to analyze the dynamics of content generation in knowledge markets. Amongst these, the Cobb-Douglas model best explains empirical data and provides an intuitive explanation for content generation through the concepts of elasticity and diminishing returns. Content generation depends on user participation and also on how specific types of content (e.g. answers) depends on other types (e.g. questions). We show that these factors of content generation have constant elasticity---a percentage increase in any of the inputs leads to a constant percentage increase in the output. Furthermore, markets exhibit diminishing returns---the marginal output decreases as the input is incrementally increased. Knowledge markets also vary on their returns to scale---the increase in output resulting from a proportionate increase in all inputs. Importantly, many knowledge markets exhibit diseconomies of scale---measures of market health (e.g., the percentage of questions with an accepted answer) decrease as a function of the number of participants. The implications of our work are two-fold: site operators ought to design incentives as a function of system size (number of participants); the market lens should shed insight into complex dependencies amongst different content types and participant actions in general social networks. 
\end{abstract}

%% file: 1_introduction.tex
\section{Introduction}

In this paper, we analyze a large group of community question answering (CQA) websites on \CQA{StackExchange} network through the Economic lens of a market. Framing \CQA{StackExchange} sites as knowledge markets has intuitive appeal: in a hypothetical knowledge market, if no one wants to answer questions, but only ask, or conversely, there are individuals who want to only answer but not ask questions, the ``market'' will collapse. What, then, is the required relationship among actions (say between questions and answers) in such a knowledge market for us to deem it healthy? Are larger markets with more participants healthier since there will be more people to ask and answer questions? 

Studying CQA websites through an economic lens allows site operators to reason about whether they should grow the user base. Since most of the popular CQA websites (e.g.\ \CQA{Quora}, \CQA{StackExchange}) do not charge participants, but instead depend on site advertisements for revenue, there is a natural temptation for operators of these sites to grow the user base so that there is increase in revenue. As we show in this paper, for most \CQA{StackExchange} sites, growth in the user base is counter-productive in the sense that they turn unhealthy---specifically, more questions remain unanswered.

Explaining the macroscopic behavior of knowledge markets is important, yet challenging. One can regress some variable of interest (say number of questions) on variables including number of users, time spent in the website among others. However, explaining why the regression curve looks the way it does is hard. As we show in this work, using an economic lens of a market allows us to model dependencies between number of participants and the amount of content, and to predict the production of content.

Our main contribution is to model CQA websites as knowledge markets, and to provide insight on the relationship between size and health of these markets. To this end, we develop models to capture content generation dynamics in knowledge markets. We analyze a set of basis functions (the functional form of how an input contributes to output) and interaction mechanisms (how the inputs interact with each other), and identify the optimal \emph{power basis} function and the \emph{interactive essential} interaction form using a prediction task on the outputs (questions, answers, and comments). This form is the well-known Cobb-Douglas form that connects production inputs with output. Using the best model fits for each \CQA{StackExchange} site, we show that the Cobb-Douglas model predicts the production of content with high accuracy.

The Cobb-Douglas function provides an intuitive explanation for content generation in \CQA{StackExchange} markets. It demonstrates that, in \CQA{StackExchange} markets, \begin{enumerate*}
  \item factors such as user participation and content dependency have \emph{constant elasticity}---percentage increase in any of these inputs will have constant percentage increase in output;
  \item in many markets, factors exhibit \emph{diminishing returns}---decrease in the marginal (incremental) output (e.g., answer production) as an input (e.g. number of people who answer) is incrementally increased, keeping the other inputs constant;
  \item markets vary according to their \emph{returns to scale}---the increase in output resulting from a proportionate increase in all inputs; and
  \item many markets exhibit \emph{diseconomies of scale}---measures of
   health decrease as a function of overall
   system size (number of participants)
 \end{enumerate*}.

There are two reasons why we see diminishing returns in the \CQA{StackExchange} markets. First, the total activity of participants for any \CQA{StackExchange} market unsurprisingly follows a power-law pattern. What is interesting is that the power-law exponent falls with increase in size for most markets, implying that new users do not participate in the same manner as earlier users. Second, we can identify a stable core of users who actively participate for long periods of time, contributing to the market health.

Finally, we show diseconomies of scale through experiments on system size, analysis of health metrics, and user exchangeability. For most \CQA{StackExchange} markets, we see that as system size grows, the ratio of answers to questions falls below a critical point, when some questions go unanswered. Furthermore, using health metrics of the number of questions with an accepted answer, and the number of questions with at least one answer, we observe that most \CQA{StackExchange} markets decline in health with increase in size. Finally, we compare the top contributors with the bottom contributors to see if they are ``exchangeable.'' Most StackExchange markets are not exchangeable in the sense the contributions of the top and the bottom contributors are qualitatively different and differ in absolute terms. These experiments on diseconomies of scale are consistent with the insight from Cobb-Douglas model of production that predicts diminishing returns.

%% file: 2_related.tex
\section{Related Work}
Our work draws from, and improves upon, several research threads.

\textbf{Sustainability.}~\citet{srba2016stack} conducted a case study on why StackOverflow, the largest and oldest of the sites in \CQA{StackExchange} network, is failing. They shed some insights into knowledge market failure such as novice and negligent users generating low quality content perpetuating the decline of the market. However, they do not provide a systematic way to understand and prevent failures in these markets.~\citet{wu2016} introduced a framework for understanding the user strategies in a knowledge market---revealing the importance of diverse user strategies for sustainable markets. In this paper, we present an alternative model that provides many interesting insights including knowledge market sustainability.

\textbf{Activity Dynamics.}~\citet{walk2016} modeled user-level activity dynamics in \CQA{StackExchange} using two factors: intrinsic activity decay, and positive peer influence. However, the model proposed there does not reveal the collective platform dynamics, and the eventual success or failure of a platform.~\citet{abufouda2017} developed two models for predicting the interaction decay of community members in online social communities. Similar to~\citet{walk2016}, these models accommodate user-level dynamics, whereas we concentrate on the collective platform dynamics.~\citet{wu2011} proposed a discrete generalized beta distribution (DGBD) model that reveals several insights into the collective platform dynamics, notably the concept of a size-dependent distribution. In this paper, we improve upon the concept of a size-dependent distribution.  

\textbf{Economic Perspective.} \citet{Kumar2010} proposed an economic view of CQA platforms, where they concentrated on the growth of two types of users in a market setting: users who provide questions, and users who provide answers. In this paper, we concentrate on a subsequent problem---the ``relation'' between user growth and content generation in a knowledge market.~\citet{butler2001} proposed a resource-based theory of sustainable social structures. While they treat members as resources, like we do, our model differs in that it concentrates on a market setting, instead of a network setting, and takes the complex content dependency of the platform into consideration. Furthermore, our model provides a systematic way to understand successes and failures of knowledge markets, which none of these models provide.  

\textbf{Scale Study.}~\citet{lin2017} examined Reddit communities to characterize the effect of user growth in voting patterns, linguistic patterns, and community network patterns. Their study reveals that these patterns do not change much after a massive growth in the size of the user community.~\citet{tausczik2017} investigated the effects of crowd size on solution quality in StackExchange communities. Their study uncovers three distinct levels of group size in the crowd that affect solution quality: topic audience size, question audience size, and number of contributors. In this paper, we examine the consequence of scale on knowledge markets from a different perspective by using a set of health metrics.

\textbf{Stability.} Successes and failures of platforms have been studied from the perspective of user retention and stability~\cite{patil2013, garcia2013, kapoor2014, ellis2016}. Notably,~\citet{patil2013} studied the dynamics of group stability based on the average increase or decrease in member growth. Our paper examines stability in a different manner---namely, by considering the relative exchangeability of users as a function of scale.

\textbf{User Growth.} Successes and failures of user communities have also been widely studied from the perspective of user growth~\cite{Kumar2006, Backstrom2006, kairam2012, Ribeiro2014, zang2016}.~\citet{kairam2012} examined diffusion and non-diffusion growth to design models that predict the longevity of social groups.~\citet{Ribeiro2014} proposed a daily active user prediction model which classifies membership based websites as sustainable and unsustainable. While this perspective is important, we argue that studying the successes and failures of communities based on content production can perhaps be more meaningful~\cite{kraut2014, zhu2014, zhu2014niche}.

\textbf{Modeling CQA Websites.} There is a rich body of work that extensively analyzed CQA websites~\cite{Adamic2008, chen2010, anderson2012, wang2013, srba2016}, along with user behavior~\cite{zhang2007, liu2011, pal2012, hanrahan2012, upadhyay2017}, roles~\cite{furtado2013, kumar2016}, and content generation~\cite{baezaYates2015, Yang2015, ferrara2017}. Notably,~\citet{Yang2015} noted the \emph{scalability problem} of CQA---namely, the volume of questions eventually subsumes the capacity of the answerers within the community. Understanding and modeling this phenomenon is one of the goals of this paper.

%% file: 3_problem.tex
\section{Problem Formulation} 
The goal of this paper is to develop a model for content generation in knowledge markets. Content is integral to the success and failure of a knowledge market. Therefore, we aim to better understand the content generation dynamics.

A model for content dynamics should have the following properties: macro-scale, explanatory, predictive, minimalistic, comprehensive.

\emph{Macro-scale:} The model should capture content generation dynamics via aggregate measures. Aggregate measures help us understand the collective market by summarizing a complex array of information about individuals, which is especially important for policy-making.

\emph{Explanatory:} The model should be insightful about the behavior of a knowledge market. Understanding market behavior is a crucial first step in designing policies to maintain a resilient, sustainable market.

\emph{Predictive:} The model should allow us to make predictions about future content generation and resultant success or failure. These market predictions are integral to the prevention and mitigation of market failures.

\emph{Minimalistic:} The model should have as few parameters as necessary, and still closely reflect the observed reality.

\emph{Comprehensive:} The model should encompass content generation dynamics for different content types (e.g., question, answer, comment) in varieties of knowledge markets. This is important for developing a systematic way to understand the successes and failures of knowledge markets.

In remaining sections we propose models that meet the aforementioned requirements, and show that our best-fit model accurately reflects the content generation dynamics and resultant successes and failures of real-world knowledge markets.

%% file: 4_model.tex
\section{Modeling Knowledge Markets}
In this section, we introduce economic production models to capture content generation dynamics in real-world knowledge markets. We first draw an analogy between economic production and content generation, and report the content generation factors in knowledge markets (Section 4.1). Then, we concentrate on the knowledge markets in \CQA{StackExchange}---presenting production models for different content types (Section 4.2).

\subsection{Preliminaries} 
Economic production mechanisms well describe content generation in knowledge markets. In economics, \emph{production} is defined as the process by which human labor is applied, usually with the help of tools and other forms of capital, to produce useful goods or services---the \emph{output}~\cite{stanford2008economics}. We assert that participants of a knowledge market function as labor to generate content such as questions and answers. Analogous to economic output, content contributes to participant utility. 

Motivated by the production analogy, we design macroeconomic production models to capture content generation dynamics in knowledge markets. In these models, instead of directly modeling content generation as a dynamic process (function of time), we model it in terms of associated factors which are dynamic. 

There are two key factors that affect content generation in knowledge markets, namely user participation and content dependency. User participation is the most important factor in deciding the quantity of generated content. The participation of more users induce more questions, answers, and other contents in a knowledge market. Content dependency also affects the quantity of generated content for different types. Content dependency refers to the dependency of one type of content (e.g., answers) on other type of content (e.g., questions). In absence of questions, there will be no answers in a knowledge market, even in the presence of many potential participants who are willing to answer. 

\subsection{Modeling \CQA{StackExchange}}
\CQA{StackExchange} is a network of community question answering websites where each site is based on a focused topic. Each user of the \CQA{StackExchange} network participates in one or more of these sites based on their interests. \CQA{StackExchange} sites are free knowledge markets where participants generate content for non-monetary reputation-based incentives. These markets are diverse, varying in theme (subject matter), size (number of users and amount of activity), and age (number of days in existence). 

We design production models for three primary content types in \CQA{StackExchange}: questions (the root content), answers (which nest below questions), and comments (which can nest either beneath questions or answers). Based on the content dependency and user roles in content generation, we propose the following relationships for question, answer and comment generation in \CQA{StackExchange} (see Table~\ref{tab:notations} for notation).

\begin{table}[thb]
	\caption{Notation used in the model}
    \label{tab:notations}
	\begin{center}    
	\begin{tabular}{cl}
	\toprule Symbol & Definition\\ \midrule
	$U_q(t)$ & \# of users who asked questions at time $t$\\ 
	$U_a(t)$ & \# of users who answered questions at time $t$\\
	$U_c(t)$ & \# of users who made comments at time $t$\\
	$N_q(t)$ & \# of active questions at time $t$\\
	$N_a(t)$ & \# of answers to active questions at time $t$\\
	$N_c^q(t)$ & \# of comments to active questions at time $t$\\
	$N_c^a(t)$ & \# of comments to active answers at time $t$\\
    $N_c(t)$ & \# of comments to active questions/answers at time $t$\\
	$f_x$ & The functional relationship for content type $x$\\ \bottomrule
	\end{tabular}
    \end{center}
\end{table}

There is a single factor in generating $N_q$ questions: the number of users $U_q$ who ask questions (askers).
\begin{equation*}
N_q = f_q(U_q)
\end{equation*}

There are two key factors in generating $N_a$ answers: the number of questions $N_q$, and the number of users $U_a$ who answer questions (answerers). 
\begin{equation*}
N_a = f_a(N_q, U_a)
\end{equation*}

There are two types of comments: comments on questions, and comments on answers. Accordingly, there are three key factors in generating $N_c$ comments: the number of questions $N_q$, the number of answers $N_a$, and the number of users $U_c$ who make comments (commenters). 
\begin{equation*}
N_c^q = f_{c^q}(N_q, U_c)
\end{equation*}
\begin{equation*}
N_c^a = f_{c^a}(N_a, U_c)
\end{equation*}
\begin{equation*}
N_c = N_c^q + N_c^a
\end{equation*}

The aforementioned relationships imply that the amount of generated content of each type depends on the function describing its factor dependent growth, and the availability of factor(s). These relationships make three assumptions. First, different content types interact only through their use of factors. Second, the functional relationships depend on the consumption or usage of each factor. Third, the functional relationships depend on the interaction among the factors---how the factors of a particular content type interact. 

Now, we transform the functional relationships into production models by first choosing a basis function to capture how a content type consumes its factor(s), and then choosing an interaction type to capture the interaction among factors.

\textbf{Basis Function.} We use a basis function to capture the effect of a given factor on a particular content type. While there is a variety of basis functions available for regression, we consider three basis functions widely used in economics and growth modeling~\cite{fekedulegn1999}: power--- $g(x) = ax^{\lambda}$; exponential--- $g(x) = ab^x$; and sigmoid--- $g(x) = \frac{L}{1+e^{k(x-x_0)}}$. 

\textbf{Interaction among the Factors.} We use an aggregate function to capture the interaction among multiple factors of a given content type. Specifically, we consider the pairwise interaction functions listed in Table~\ref{tab:interaction}. 

\begin{table}[htb]
  \caption{Pairwise interaction between factors with contour}
  \label{tab:interaction}
  \centering
  \begin{tabular}{m{.28\textwidth}c}
	\vspace{-5pt}
    \uline{Essential:} Essential factors are both required for content generation, with zero marginal return for a single factor. For a pair of essential factors, content generation is determined by the more limiting factor: $z = min(y_1, y_2)$~\cite{tilman1980}. This is known as Liebig's law of the minimum.
    &
    \begin{minipage}{.17\textwidth}
      \includegraphics[width=\textwidth, height=\textwidth]{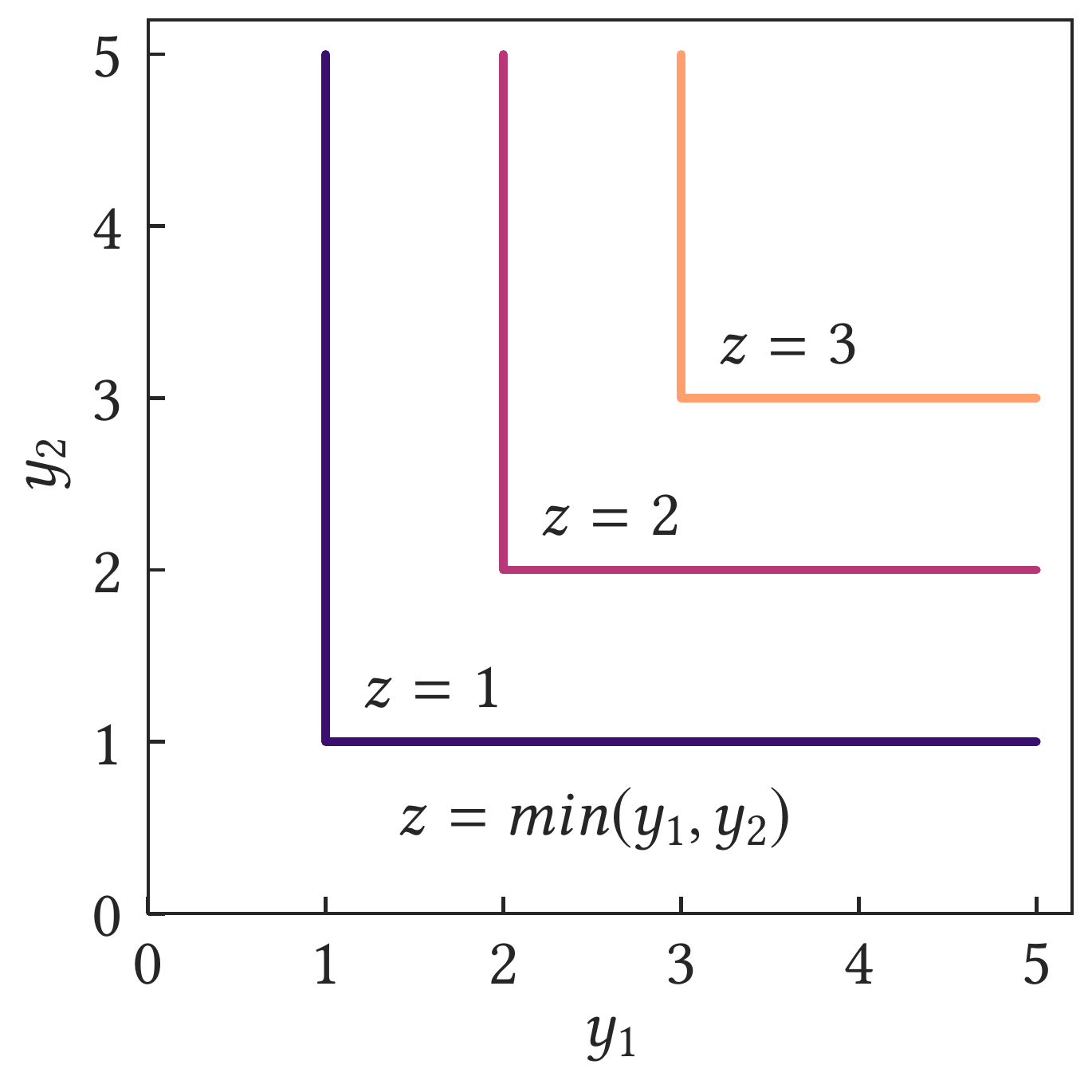}
    \end{minipage}
    \\ 
    \vspace{-5pt}
    \uline{Interactive Essential:} In interactive essential interaction, we get diminishing return (instead of zero return) for a single factor: $z = y_1y_2$~\cite{tilman1980}. If factors are consumed using power basis function, i.e., $y_i = ax^\lambda_i$, it captures Cobb-Douglas production function.
    &
    \begin{minipage}{.17\textwidth}
      \includegraphics[width=\textwidth, height=.975\textwidth]{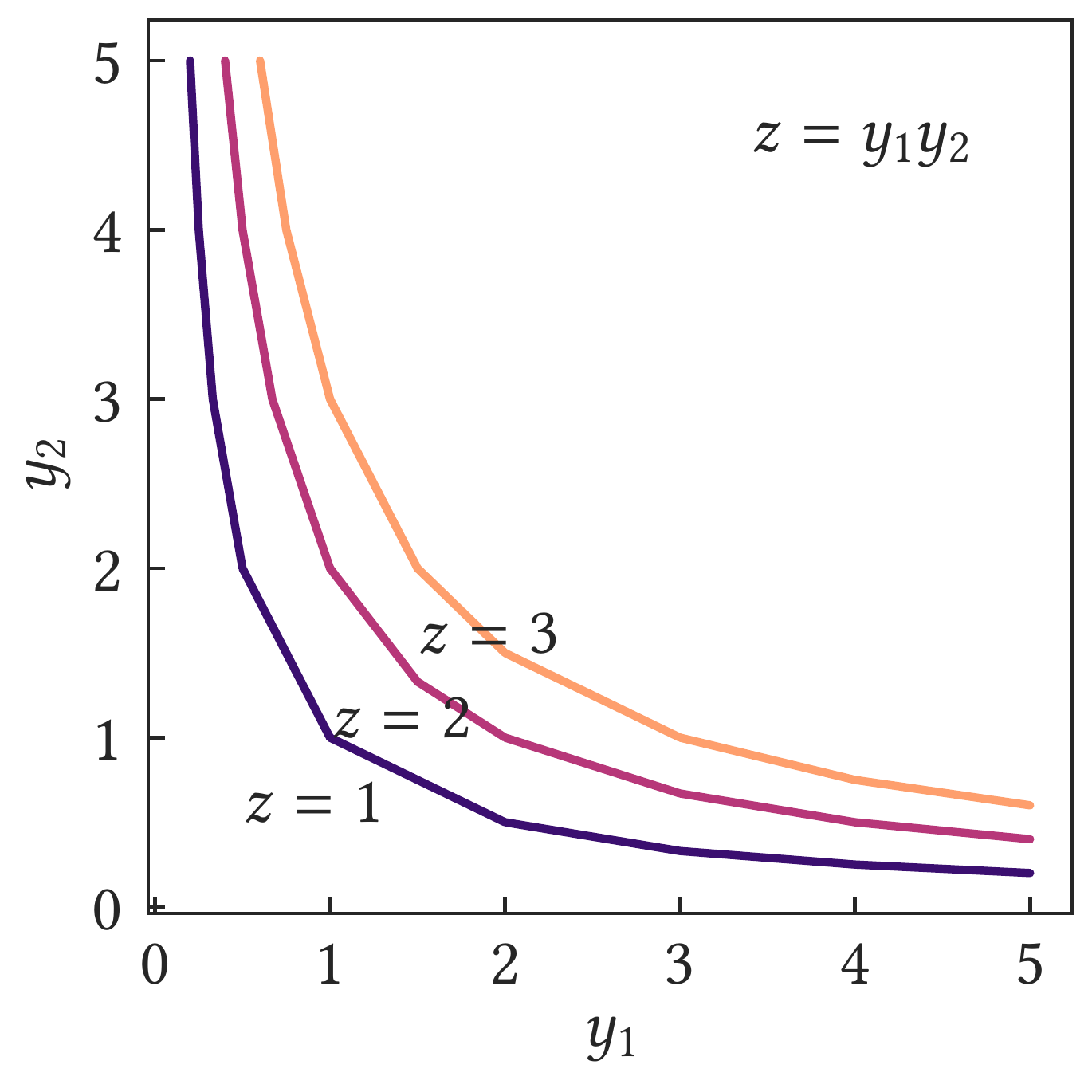}
    \end{minipage}
    \\
    \vspace{-5pt}
    \uline{Antagonistic:} For antagonistic factors, content generation is determined solely by the availability of the factor which yields the largest return: $z = max(y_1, y_2)$~\cite{tilman1980}. This interaction implies that the production process has maximum possible efficiency. 
    &
    \begin{minipage}{.17\textwidth}
      \includegraphics[width=\textwidth, height=.975\textwidth]{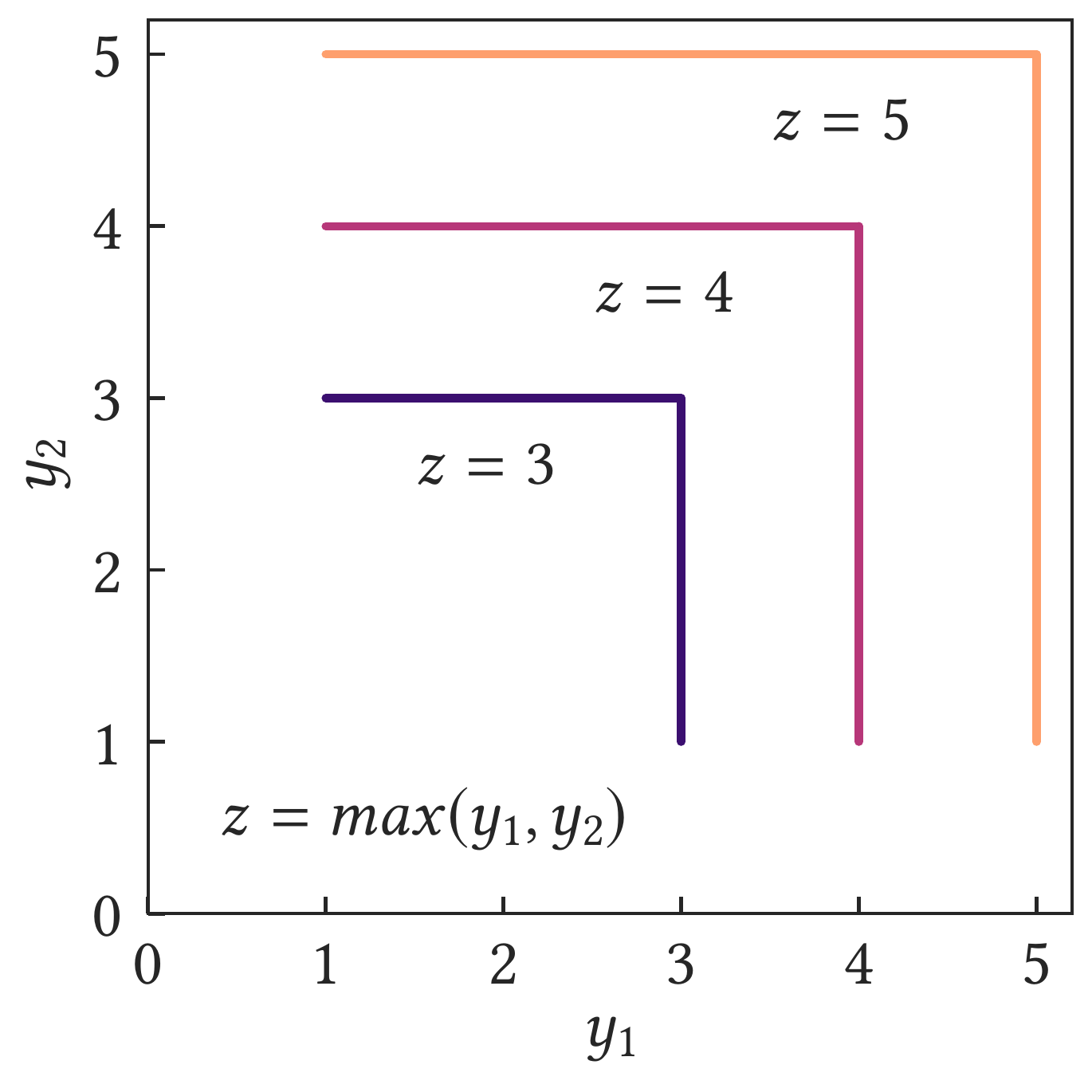}
    \end{minipage}
    \\
    \vspace{-5pt}
    \uline{Substitutable:} Factors that can each support production on their own are substitutable relative to each other: $z = w_1y_1 + w_2y_2$~\cite{tilman1980}. This implies that there exists some equivalence between the two factors. This is analogous to the general additive models.
    &
    \begin{minipage}{.17\textwidth}
      \includegraphics[width=\textwidth, height=.975\textwidth]{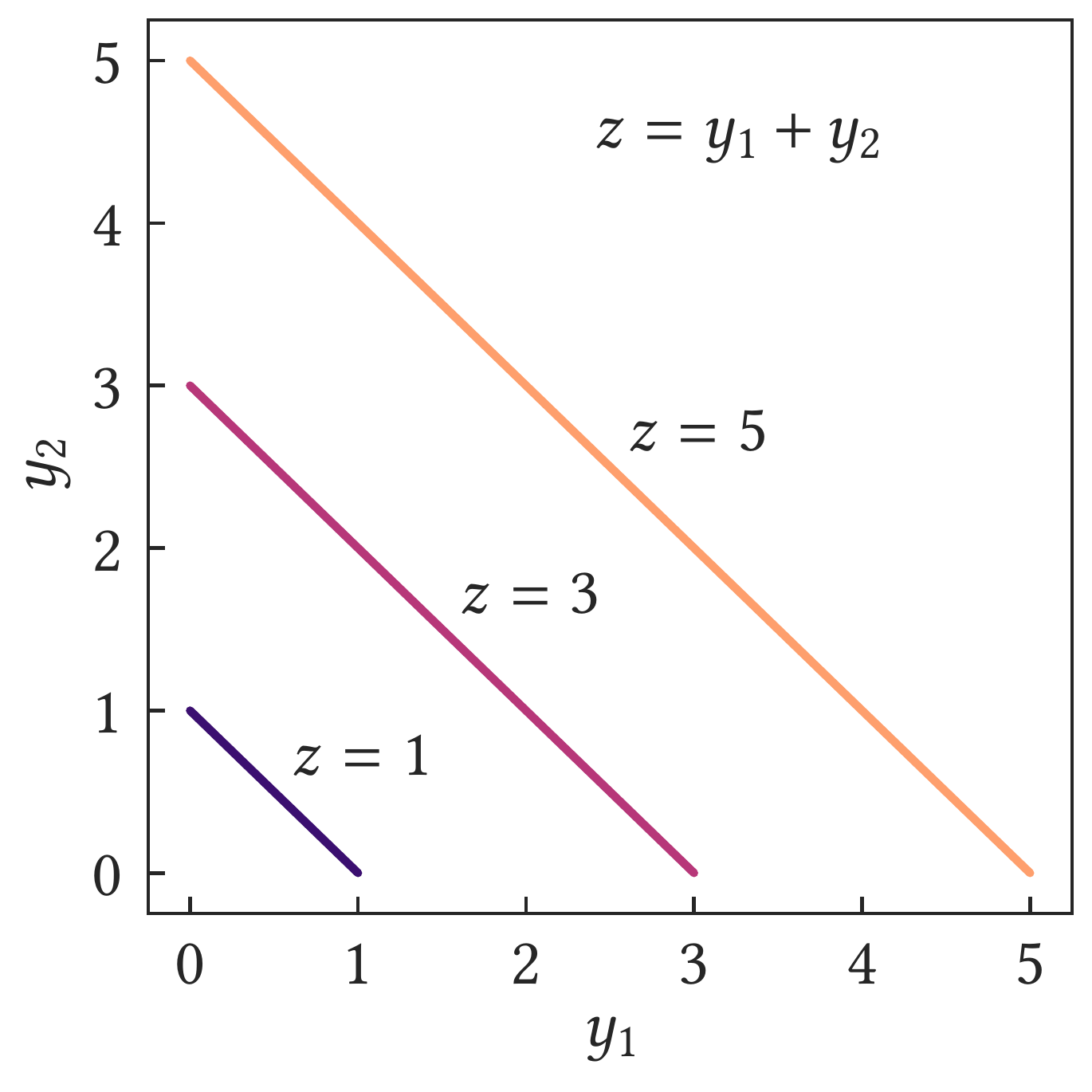}
    \end{minipage}
    \\
  \end{tabular}
\end{table}

We combine a basis function and an interaction type to design production models for different content types. For example, answer generation can be modeled using power basis and essential interaction as $N_a = min(a_1N_q^{\lambda_1}, a_2U_a^{\lambda_2})$. We consider twelve such possible models (combination of three basis and four interaction type) for answer and comment generation in StackExchange.

\textbf{User Role Distribution.} A fundamental assumption of our model is the awareness of user roles (e.g., asker, answerer, and commenter) and their distribution (e.g., how many users are askers?). We empirically observe that all StackExchange markets have a stable distribution of user roles. In fact, given the number of users, we can accurately predict the number of participants for each role. 

\begin{figure}[hbt]
\centering
\includegraphics[scale=0.45]{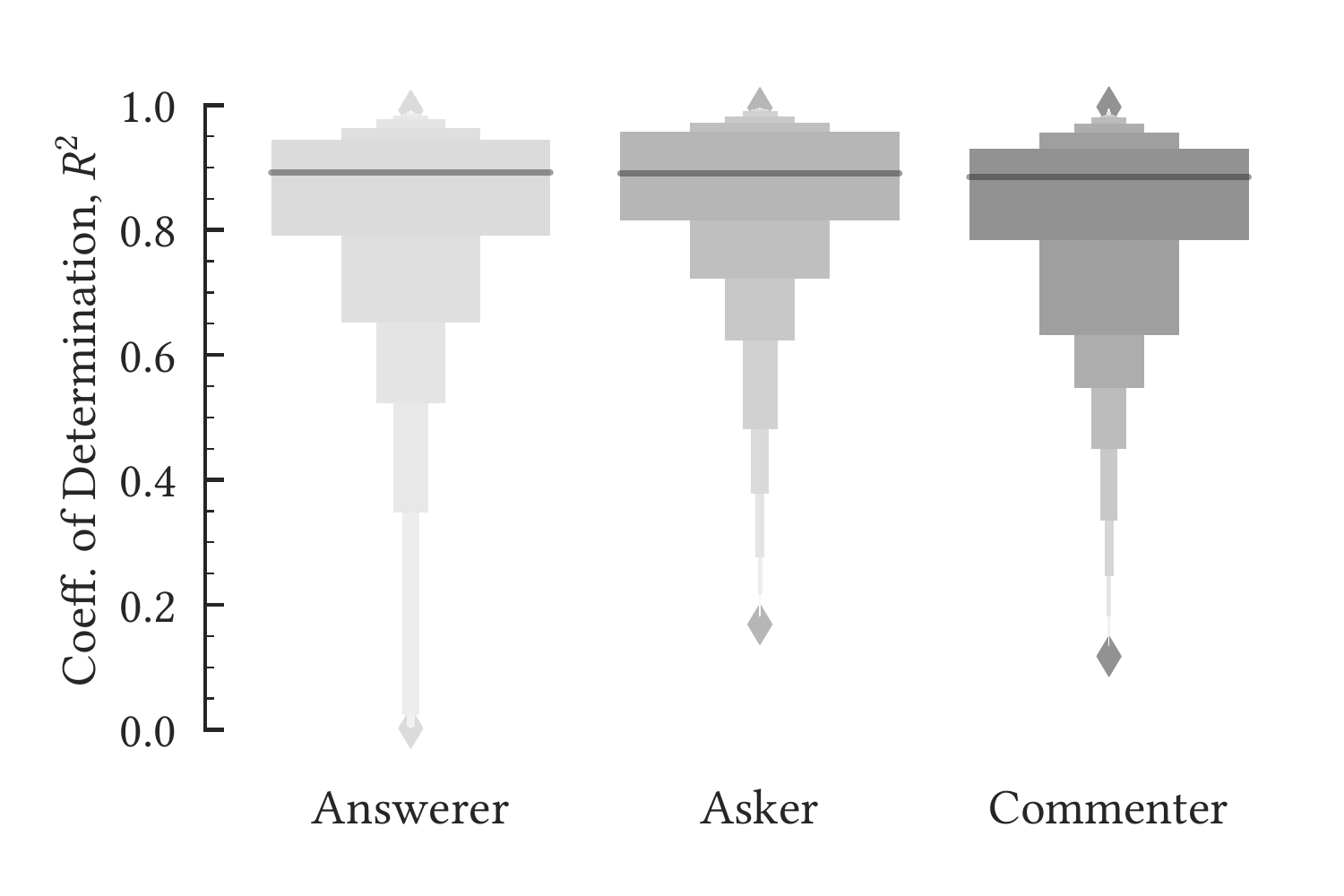}
\caption{The distribution of coefficient of determination, $R^2$, for regressing user roles across 156 StackExchange. In most StackExchange, the role distribution is stable---as manifested by the arrangement of $R^2$ in letter-value plots.}
\label{fig:roles}
\end{figure}

We apply linear regression to determine the number of participants $U_x$ of a particular role $x \in \{q, a, c\}$ from the number of users $U$ in a StackExchange market. For each market, we compute three distinct coefficients of determination, $R^2$, for predicting three roles (asker, answerer, and commenter) using linear regression. In Figure~\ref{fig:roles} we show the distribution of $R^2$ for regressing user roles across 156 StackExchange markets. We use letter value plots\footnote{The letter-value plot display information about the distribution of a variable~\cite{Hofmann2017}. It conveys precise estimates of tail behavior using letter values; boxplots lack such precise estimation.} to present these distributions---showing precise estimates of their tail behavior. We observe that, in most markets, the $R^2$ values are close to 1. Further, the tail capturing low $R^2$ values consists of markets with a relatively small number of monthly users.

\textbf{Number of Users.} The number of users is the only free input to our content generation models; the remaining inputs are functions of the number of users. In all these models, the growth or decline of number of users is exogenous---determined outside the model, by non-economic forces.

%% file: 5_dataset.tex
\section{Dataset} 
We collected the latest release (September, 2017) of the \CQA{StackExchange} dataset. This snapshot is a complete archive of all activities in \CQA{StackExchange} sites. There are 169 sites in our collected dataset. For the purpose of empirical analysis, we only consider the sites that have been active for at least 12 months beyond the \emph{ramp up} period (site created, but few or no activity). There are 156 such sites. The age of these sites vary from 14 months to 111 months, number of users from 1,072 to 547,175, and the number of posts (questions and answers) from 1,600 to 1,985,869. Further, the sites have small overlaps in user base; therefore, we can reasonably argue that the underlying markets are independent. 

\begin{figure}[hbt]
\centering
\includegraphics[scale=0.45]{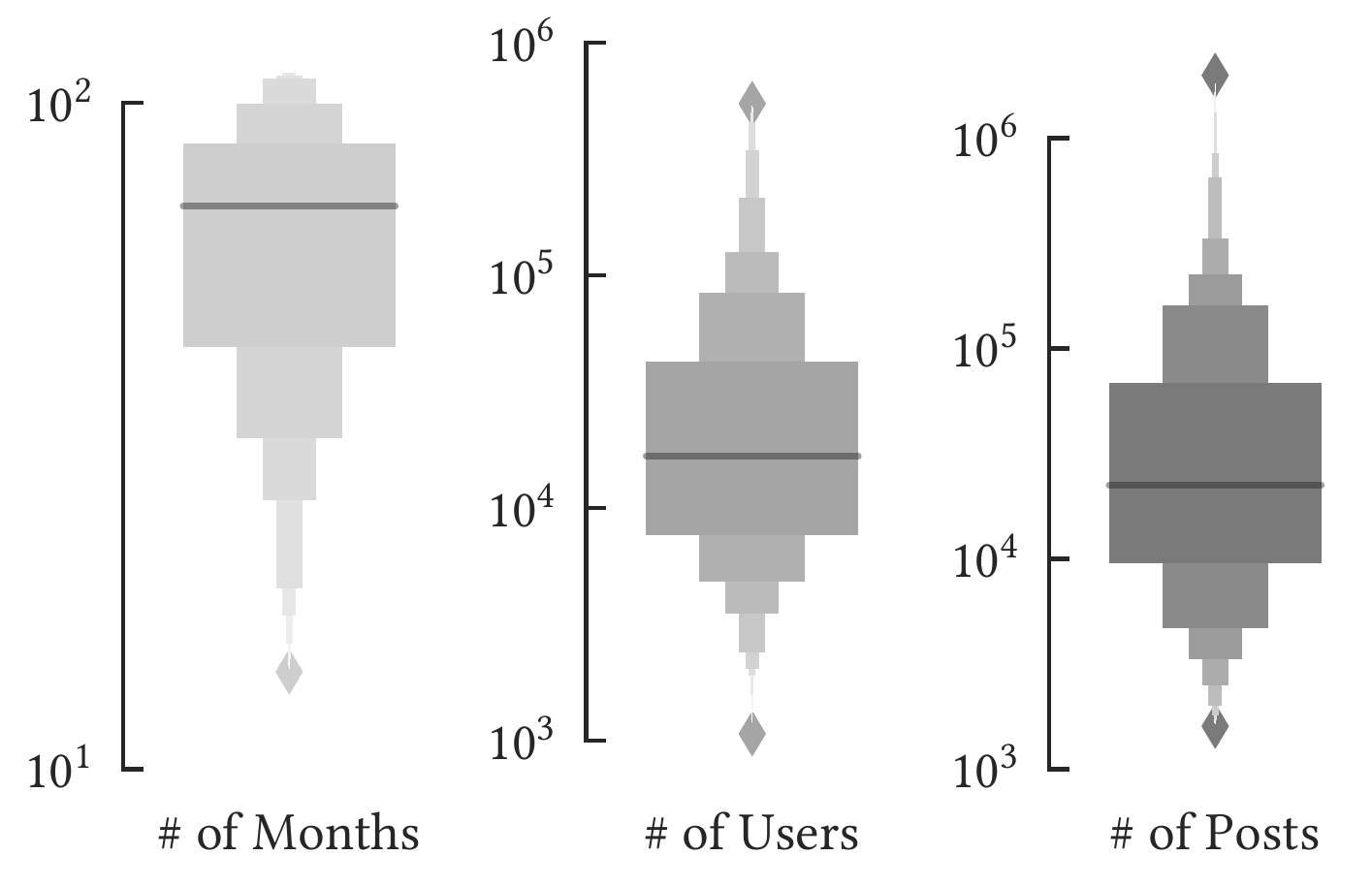}
\caption{The log-scale distribution of number of months (age), number of users, and number of posts for 156 \CQA{StackExchange} sites.}
\label{fig:dataset}
\end{figure}

In Figure~\ref{fig:dataset} we present letter value plots (in log-scale) to show the distribution of number of months (age), number of users, and number of posts for 156 \CQA{StackExchange} sites. 

%% file: 6_empirical.tex
\section{Evaluating Our Proposed Models}
In this section, we identify optimal models (basis and interaction) based on the accuracy of fitting content generation time series observed in our dataset (Section 6.1), and evaluate the performance of the optimal models in predicting content volume in long run (Section 6.2).

\subsection{Model Fitting}
We fit each variant of production model (basis and interaction), for each content type, to the observed content generation time series (monthly granularity), in each \CQA{StackExchange} site. Notice that among the different variants of production models, the models using power or exponential basis have a parsimonious set of parameters. For example, answer generation model using power basis function requires only three parameters for interactive essential interaction (See Section 4.2), and four parameters for remaining interaction types. In contrast, answer generation model using sigmoid basis function requires five parameters for interactive essential interaction, and six parameters for remaining interaction types. 

\textbf{Parameter Estimation.} We learn the best-fit parameters for capturing the observed content generation time series. We restrict some parameters of our production models to be non-negative, e.g., non-negative exponents in power basis. These restrictions are important because the underlying factors positively affect the output. We use the trust-region reflective algorithm \cite{branch1999} to solve our constrained least square optimization problem. The algorithm is appropriate for solving non-linear least squares problems with constraints.

\textbf{Evaluation Method.} We evaluate fitting accuracy using four metrics: root mean square error (RMSE), normalized root mean square error (NRMSE), explained variance score (EVS), and Akaike information criterion (AIC). Given two series for each content type, the observed series $N(t)$, and the prediction $\hat{N(t)}$ of the series by a model with $k$ parameters, we compute the four metrics as follows:\\ RMSE = $\sqrt{\frac{1}{T}\sum_{t=1}^{T}(N(t)-\hat{N(t)})^2}$; NRMSE = $\frac{RMSE}{max(N(t))-min(N(t))}$; EVS = $1-\frac{Var(N(t)-\hat{N(t)})}{Var(N(t))}$; AIC = $T*ln(\frac{1}{T}\sum_{t=1}^{T}(N(t)-\hat{N(t)})^2)+2k$. Among the four metrics, RMSE and NRMSE are error metrics (low value implies good fit), AIC is an information theoretic metric to capture the trade-off between model complexity and goodness-of-fit (low value implies good model), and EVS refers to a model's ability to capture variance in data (high value implies good model).

\textbf{Fitting Results.} We compare the fitting accuracy of production models for all \CQA{StackExchange} sites using the four metrics. Each metric is summarized via the mean, across all sites, for each content type. We use content generation time series with monthly granularity as observed data. We found that the models with the exponential and sigmoid basis functions do not fit the data for many \CQA{StackExchange} sites. Accordingly, in Table~\ref{tbl:model_fit} we only present the results for production models with the power basis and different interaction types. Notice that the models with interactive essential interaction outperform the remaining models for all metrics and content types. We performed paired $t$-tests to determine if the improvements for interactive essential interaction are statistically significant; the results are positive with $p<0.01$.

\begin{table}[ht]
	\caption{The comparison of fitting accuracy of production models (with power basis and different interaction types) for all \CQA{StackExchange} sites. The models with interactive essential interaction outperform the remaining models for all metrics and content types. The improvements for interactive essential interaction are statistically significant, validated via paired t-tests, where $p<0.01$.}
	\label{tbl:model_fit}
	\begin{center}
	\begin{tabular}{llcccc}
    \toprule
    \multirow{2}{*}{Content} & Interaction & Avg. & Avg. & Avg. & Avg.\\
    & Type & RMSE & NRMSE & EVS & AIC\\
    \midrule
    Question & Single Factor & 25.74 & 0.09 & 0.79 & 104.47\\
    \midrule
    \multirow{4}{*}{Answer} & Essential & 70.31 & 0.09 & 0.79 & 208.82\\
    & I. Essential & \textbf{64.62} & \textbf{0.08} & \textbf{0.83} & \textbf{196.39}\\
    & Antagonistic & 72.77 & 0.09 &  0.78 & 210.96\\
    & Substitutable & 68.90 & 0.09 & 0.81 & 207.61\\
    \midrule
    \multirow{4}{*}{Comment} & Essential & 146.64 & 0.08 & 0.83 & 328.25\\
    & I. Essential & \textbf{137.23} & \textbf{0.08} & \textbf{0.85} & \textbf{318.24}\\
    & Antagonistic & 155.97 & 0.09 &  0.82 & 334.12\\
    & Substitutable & 155.43 & 0.09 & 0.82 & 335.10\\
    \bottomrule
	\end{tabular}
	\end{center}
\end{table}

Thus we use production models with power basis and interactive essential interaction for prediction tasks.

\subsection{Forecasting Content Generation} 
We apply production models with power basis and interactive essential interaction to forecast content volume in long run---one year ahead in the future. Specifically, we train each model using the content generation data from the first 12 months (beyond the ramp period), and then examine how well the model forecasts content dynamics in the next 12 months. We validate the forecasting capability by examining the overall prediction error (NRMSE). 

We compute the prediction NRMSE across all StackExchange sites, and summarize the results using the mean ($\mu$) and variance ($\sigma$)--- (i) question: $\mu = 0.11$, $\sigma = 0.08$; (ii) answer: $\mu = 0.12$, $\sigma = 0.09$; (iii) comments: $\mu = 0.11$, $\sigma = 0.10$. Notice that our models can forecast future content dynamics with high accuracy. We performed these experiments for different time granularity, e.g., week, month, quarter, and reached a consistent conclusion. We do not report these results for brevity.

%% file: 7_interpretation.tex
\section{Characterizing Knowledge Markets}
In this section, we characterize the knowledge markets in \CQA{StackExchange}. We explain the best-fit models and their foundations (Section 7.1), reveal two key distributions that control the markets (Section 7.2), and uncover the stable core that maintains market equilibrium (Section 7.3).

\subsection{Model Interpretation} 
First, we explain the best-fit models found in Section 6.1. We observe that content generation in \CQA{StackExchange} markets are best modeled through the combination of power basis and interactive essential interaction. In addition, we found that the best-fit exponents ($\lambda$ parameter in basis $g(x) = ax^\lambda$, where $x$ is a factor) of these models lie between 0 and 1 (inclusive), for all factors of all content types, for all \CQA{StackExchange} markets. 

A model that uses the power basis (where exponents lie between 0 and 1) and interactive essential interaction is known as the Cobb-Douglas production function~\cite{wiki}. In its most standard form for production of a single output $z$ with two inputs $x_1$ and $x_2$, the function is: 
$$z = ax_1^{\lambda_1}x_2^{\lambda_2}.$$
Here, the coefficient $a$ represents the \emph{total factor productivity}---the portion of output not explained by the amount of inputs used in production~\cite{wiki}. As such, its level is determined by how efficiently the inputs are utilized in production. The exponents $\lambda_i$ represent the \emph{output elasticity} of the inputs---the percentage change in output that results from the percentage change in a particular input~\cite{wiki}. 

The Cobb-Douglas function provides intuitive explanation for content generation in \CQA{StackExchange} markets. In particular, the explanation stands on three phenomena or principles: constant elasticity, diminishing returns, and returns to scale.

\textbf{Constant Elasticity.} In \CQA{StackExchange} markets, factors such as user participation and content dependency have \emph{constant elasticity}---percentage increase in any of these inputs will have constant percentage increase in output~\cite{wiki}, as claimed by the corresponding exponents in the model. For example, in \SE{academia} ($N_a = 6.93N_q^{0.18}U_a^{0.65}$), a 1\% increase in number of answerers ($U_a$) leads to a 0.65\% increase in number of answers ($N_a$). 

\textbf{Diminishing Returns.} For a particular factor, when the exponent is less than 1, we observe \emph{diminishing returns}---decrease in the marginal (incremental) output as an input is incrementally increased, while the other inputs are kept constant~\cite{wiki}. This ``law of diminishing returns'' has many interesting implications for the \CQA{StackExchange} markets, including the diminishing benefit of having a new participant in a market. For example, in \SE{academia}, if the number of answerers is 100, then the marginal contribution of a new answerer is $c(101^{0.65} - 100^{0.65}) = 0.129c$, where $c$ is a constant; in contrast, if the number of answerers is 110, then the marginal contribution of a new answerer is $c(111^{0.65} - 110^{0.65}) = 0.125c$. Thus, for answer generation in \SE{academia}, including a participant when the number of participants (system size) is 110 is likely to be less beneficial compared to including a participant when the system size is 100.

\textbf{Returns to scale.} The knowledge markets in \CQA{StackExchange} vary in terms of scale efficiency, as manifested by their \emph{returns to scale}---the increase in output resulting from a proportionate increase in all inputs~\cite{wiki}. If a market has high returns to scale, then greater efficiency is obtained as the market moves from small- to large-scale operations. For example, in \SE{academia}, for answer generation, the returns to scale is $0.18+0.65=0.83<1$. The market becomes less efficient as answer generation is expanded, requiring more questions and answerers to increase the number of answers by same amount. 

\subsection{Two Key Distributions} 
Next, we discuss two key distributions that control content generation in knowledge markets, namely participant activity and subject POV (perspective). These two distributions induce the three phenomena reported in section 7.1.

\textbf{Participant Activity.} The distribution of participant activities implicitly drives a market's return in terms of user participation, as manifested by the corresponding exponent. For example, in a hypothetical knowledge market where each answerer contributes equally, the answer generation model should be $N_a = AN_q^{\lambda_1}U_a^{1.0}$. In reality, the distribution of participant activities is a size dependent distribution controlled by the number of participants (system size). As the system size increases, most participants contribute to the head of the distribution (few activities), whereas very few join the tail (many activities). 

\begin{figure}[b]
\centering
\includegraphics[scale=0.39]{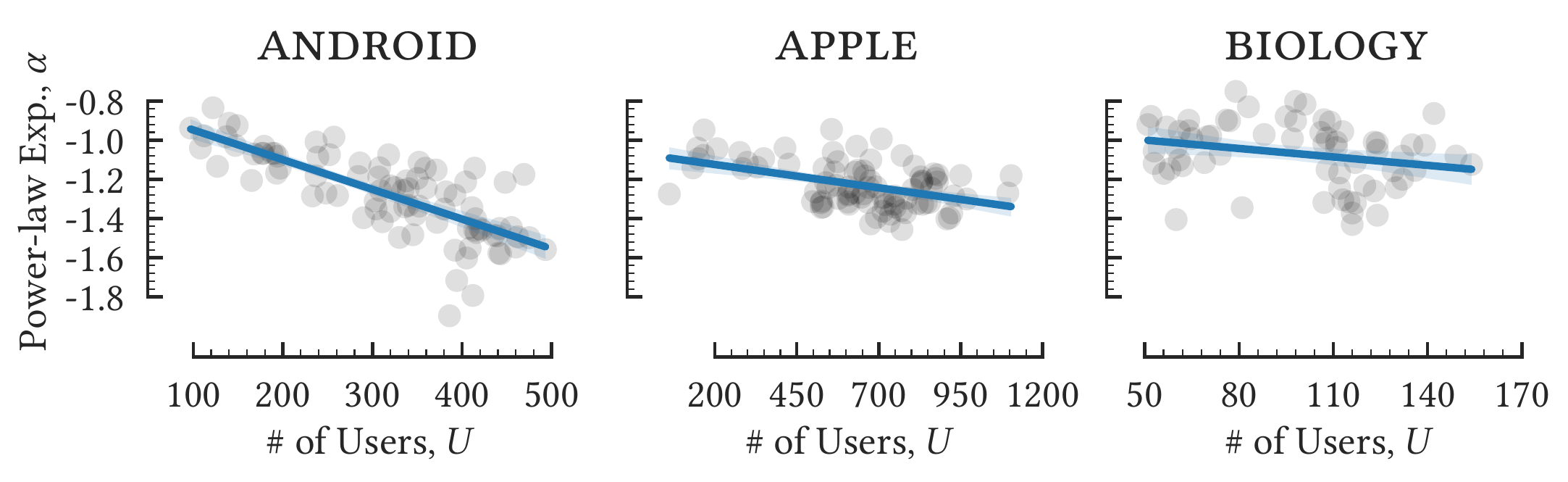}
\caption{The visibility of size dependent distribution: strong---\SE{android}; moderate---\SE{apple}; and weak---\SE{biology}. In most \CQA{StackExchange} markets, the power-law exponent decreases with system size, similar to \SE{android}. In other markets, there exists a non-zero correlation between system size and power-law exponent.}
\label{fig:sdd}
\end{figure}

We systematically reveal the size dependent distribution for participant activities in three steps. First, we empirically a fit power-law distribution to the activities of participants in a month, for each month, for each \CQA{StackExchange} market. We follow the standard procedure to fit a power-law distribution~\cite{adamic2000zipf}. We observe that the power-law well describes the monthly activity distributions. Second, we plot the exponents of the power-law against the number of participants for all observed months in a market, for each market in \CQA{StackExchange}. We observe that for most \CQA{StackExchange} markets, the power-law exponent decreases as the system size increases. Third, we apply linear regression to reveal the relationship between power-law exponent and system size. We observe that in general power-law exponent is negatively correlated with system size. This negative correlation is strongly visible in big knowledge markets that have at least 500 monthly participants in each month.

In Figure~\ref{fig:sdd} we present empirical evidence of the size dependent distribution for answer generation in three \CQA{StackExchange} markets: \SE{android}, \SE{apple}, and \SE{biology}. We choose these examples to cover three possible visibilities of the size dependent distribution, as manifested by the correlation between 
the power-law exponent and system size---strong correlation ($|r^2|\geq 0.5$), moderate correlation ($0.3\leq |r^2|<0.5$), and weak correlation ($|r^2|<0.3$).

\textbf{Subject POV.} The distribution of subject POV implicitly drives a market's return in terms of content dependency, as manifested by the corresponding exponent. Subject POV refers to the number of distinct perspectives on a particular content (e.g., questions) that imposes a conceptual limit to the number of dependent contents (e.g., answers). For example, an open-ended question such as \lq What's your favorite book?\rq\ has many possible answers, whereas a close-ended question such as \lq What's the solution for 3x+5 = 2?\rq\ has a single correct answer. In reality, most questions are neither completely open-ended nor completely closed; however, from an answerer's perspective, there's a diminishing utility in answering a question that already has an answer. This diminishing utility varies from question to question---questions asking for recommendations attract many answers, whereas questions seeking factual information attract few answers. 

\subsection{Uncovering the Stable Core} 
We uncover a stable user community in each \CQA{StackExchange} market, that maintains the \emph{dynamic equilibria}---the increase or decrease in overall user community does not affect the Cobb-Douglas models. We assert that this stable user community generates a large fraction of high-threshold contents that require more effort, e.g, answers and comments, whereas the remaining users are unstable and contribute a small fraction of high-threshold contents.

\begin{figure}[t]
\centering
\includegraphics[scale=0.45]{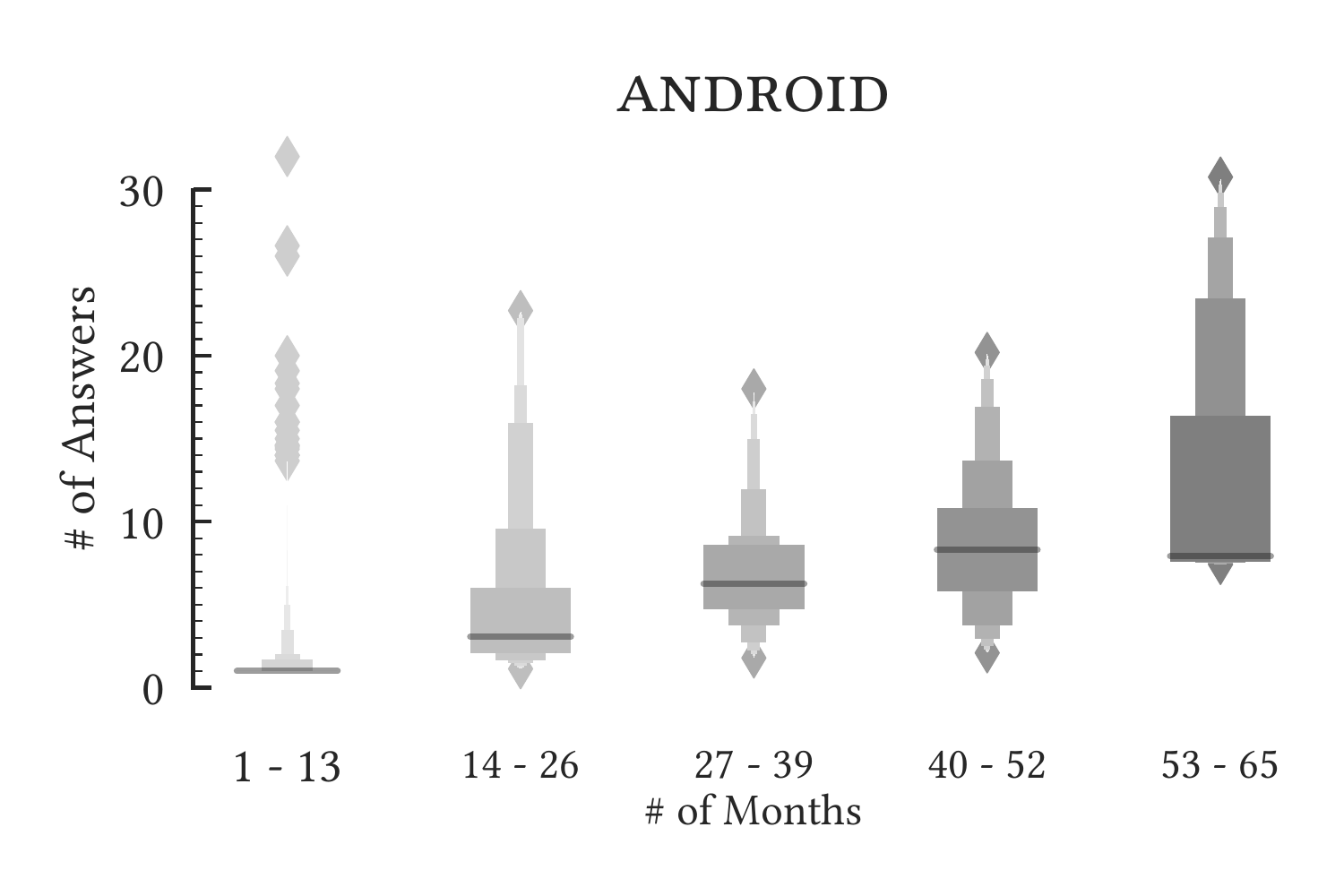}
\caption{The distribution of monthly answer contribution of users with different \# of active months, for \SE{android}. The users who contribute for many months also contribute a large number of answers.}
\label{fig:stable_core}
\end{figure}

We reveal the presence of the stable core by summarizing the answer contribution of users with different tenure levels (\# of active months). First, for each \CQA{StackExchange} market, we apply equal-width binning to categorize its users into five tenure levels. Then, we plot the distribution of monthly answer contribution by the users of each category using a letter-value plot. We present the letter-value plots for \SE{android} in Figure~\ref{fig:stable_core}. We observe that monthly answer contribution is an increasing function of tenure level---the users who contribute for many months also contribute a large number of answers.

%% file: 8_diseconomies.tex
\section{Failures at Scale}
In this section, we discuss how and why knowledge markets may fail at scale. We first empirically examine diseconomies of scale (Section 8.1), then analyze the effects of scale on market health (Section 8.2), and finally study user exchangeability under scale changes (Section 8.3).

\begin{figure}[b]
\centering
\includegraphics[scale=0.39]{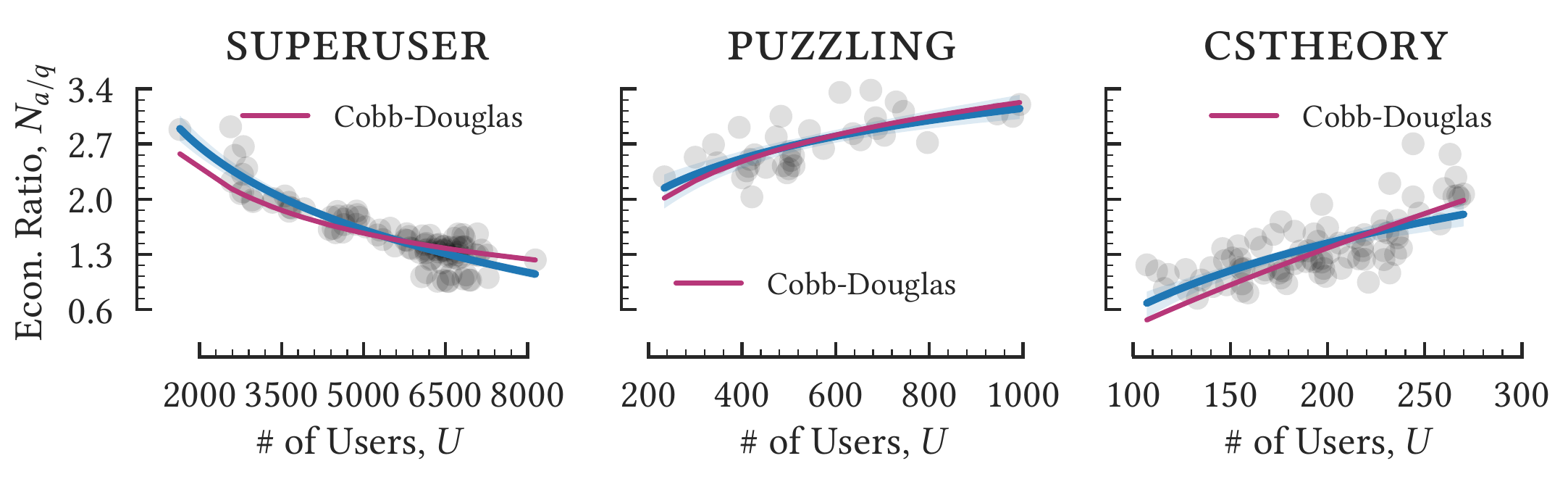}
\caption{Diseconomies/economies of scale: the ratio of answers to questions decreasing/increasing with the increase in number of users. Most \CQA{StackExchange} markets exhibit diseconomies of scale. Examples: strong diseconomies---\SE{superuser}; weak economies---\SE{puzzling}; and strong economies---\SE{cstheory}.}
\label{fig:diseconomy}
\end{figure}

\subsection{Diseconomies of Scale}
First, we examine disceconomies of scale---the ratio of answers to questions declining with the increase in number of users. The opposite of diseconomies is economies, when the ratio increases with the increase in number of users. The concept of diseconomies is important because a decrease in the answer to question ratio implies an increase in the gap between market supply (answer) and demand (question). In fact, if the ratio falls below 1.0, the gap becomes critical---guaranteeing there will be some questions with no answers. 

In Figure~\ref{fig:diseconomy} we present the economies and diseconomies of scale in three StackExchange markets: \SE{cstheory}, \SE{puzzling} and \SE{superuser}. We choose these examples to cover three cases: strong diseconomies, strong economies, and weak economies. Among the three markets, \SE{superuser} shows strong diseconomies of scale: if the number of users increases by 1\%, then the answer to question ratio declines by 0.95\%. The other two markets show economies of scale, where \SE{cstheory} shows strong economies: if the number of users increases by 1\%, then the answer to question ratio increases by 0.8\%; and \SE{puzzling} shows weak economies: if the the number of users increases by 1\%, then the answer to question ratio increases by 0.2\%. Note that most markets, especially the ones with more than 500 monthly active participants, exhibit diseconomies of scale similar to \SE{superuser}. Only five markets exhibit strong economies of scale in \CQA{StackExchange}: \SE{cstheory}, \SE{expressionengine}, \SE{puzzling}, \SE{ja\_stackoverflow}, and \SE{softwareengineering}.

The Cobb-Douglas curves well fit the empirical trends of economies and diseconomies (as shown in Figure~\ref{fig:diseconomy}). We derive these curves by dividing the answer models by the corresponding question models, and subsequently developing curves that capture economies and diseconomies ($N_{a/q}$) as a function of number of users (system size). We get similar curves via log regression. Between the two model types, the Cobb-Douglas models provide better explanation.

The Cobb-Douglas models well explain the economies and disceconomies of scale. As per the models, the primary cause of disceconomies is the difference between the diminishing returns of questions and answers for user participation. In other words, in most markets, for user input, the marginal question output is higher compared to the marginal answer output, i.e., an average user is likely to ask more questions and provide few answers. This causes the ratio of answers to questions to decline with an increase in the number of users. 

\subsection{Analyzing Health}
Next, we examine the disadvantage of scale through two health metrics: $H_1$---the fraction of answered questions (questions with at least one answer); and $H_2$---the fraction of questions with an accepted answer (questions for which the asker marked an answer as accepted). $H_1$ and $H_2$ capture the true gap between market supply (answers) and demand (questions). An increase in the number of users may cause a decline in $H_1$ and $H_2$, as both metrics are related to the ratio of answers to questions. In fact, if the ratio falls below 1.0, it guarantees the decline of both metrics. 

\begin{figure}[t]
\centering
\includegraphics[scale=0.39]{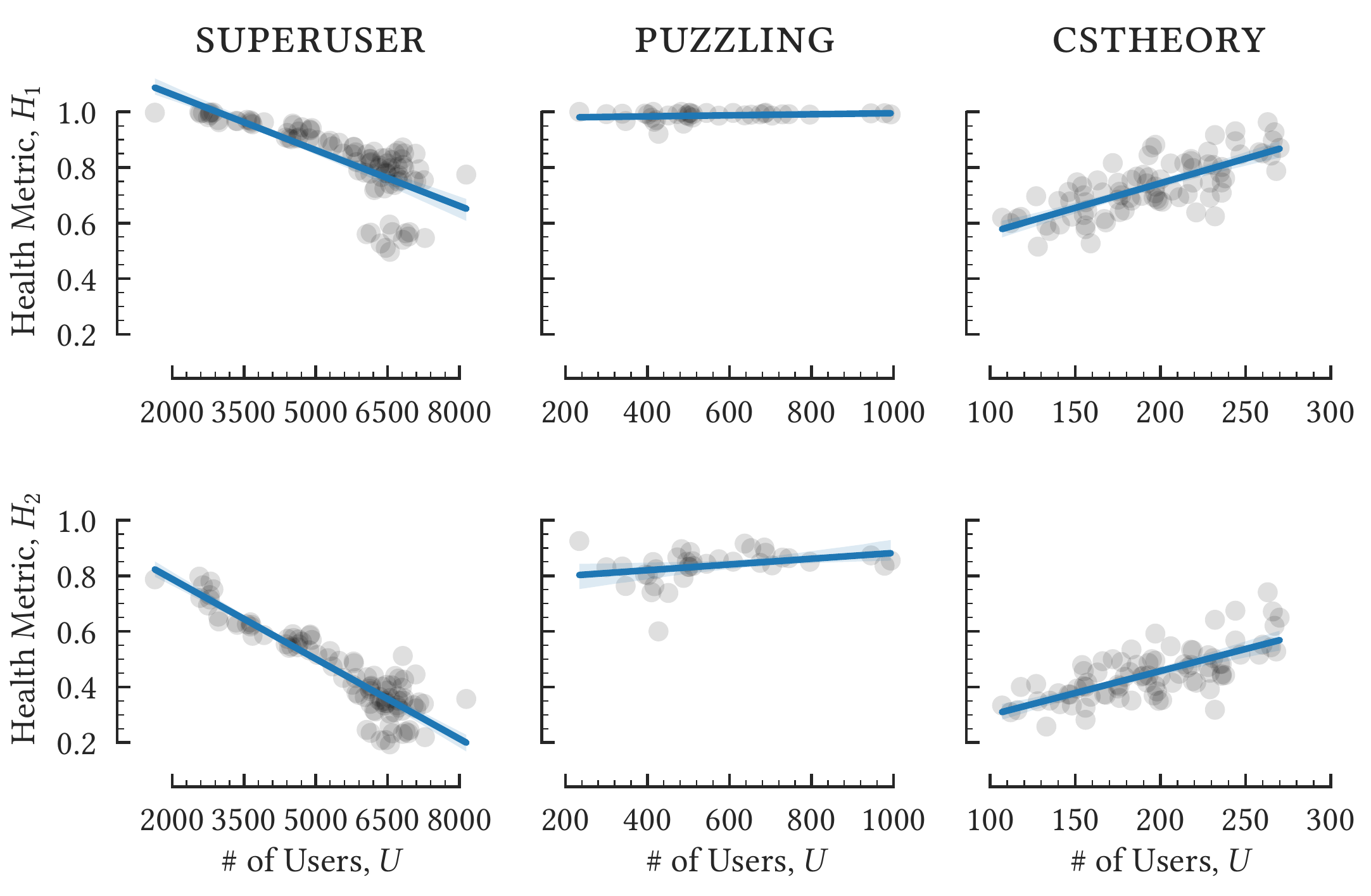}
\caption{Health disadvantage/advantage of scale: $H_1$---the fraction of answered questions, and $H_2$---the fraction of questions with accepted answer, decreasing/increasing with the increase in number of users. Most \CQA{StackExchange} markets exhibit health disadvantage at scale. Examples: disadvantage---\SE{superuser}; neutral---\SE{puzzling}; and advantage---\SE{cstheory}.}
\label{fig:health}
\end{figure}

In Figure~\ref{fig:health} we present the health advantage and disadvantage of scale (through $H_1$ and $H_2$) for three StackExchange markets: \SE{cstheory}, \SE{puzzling} and \SE{superuser}. We observe that the results are consistent with our analysis of economies and diseconomies---\SE{cstheory} exhibits health advantage at scale, \SE{puzzling} remains stable, whereas \SE{superuser} exhibits disadvantage at scale. These three examples cover the possible health effects of scale in knowledge markets. 

\begin{figure}[b]
\centering
\includegraphics[scale=0.39]{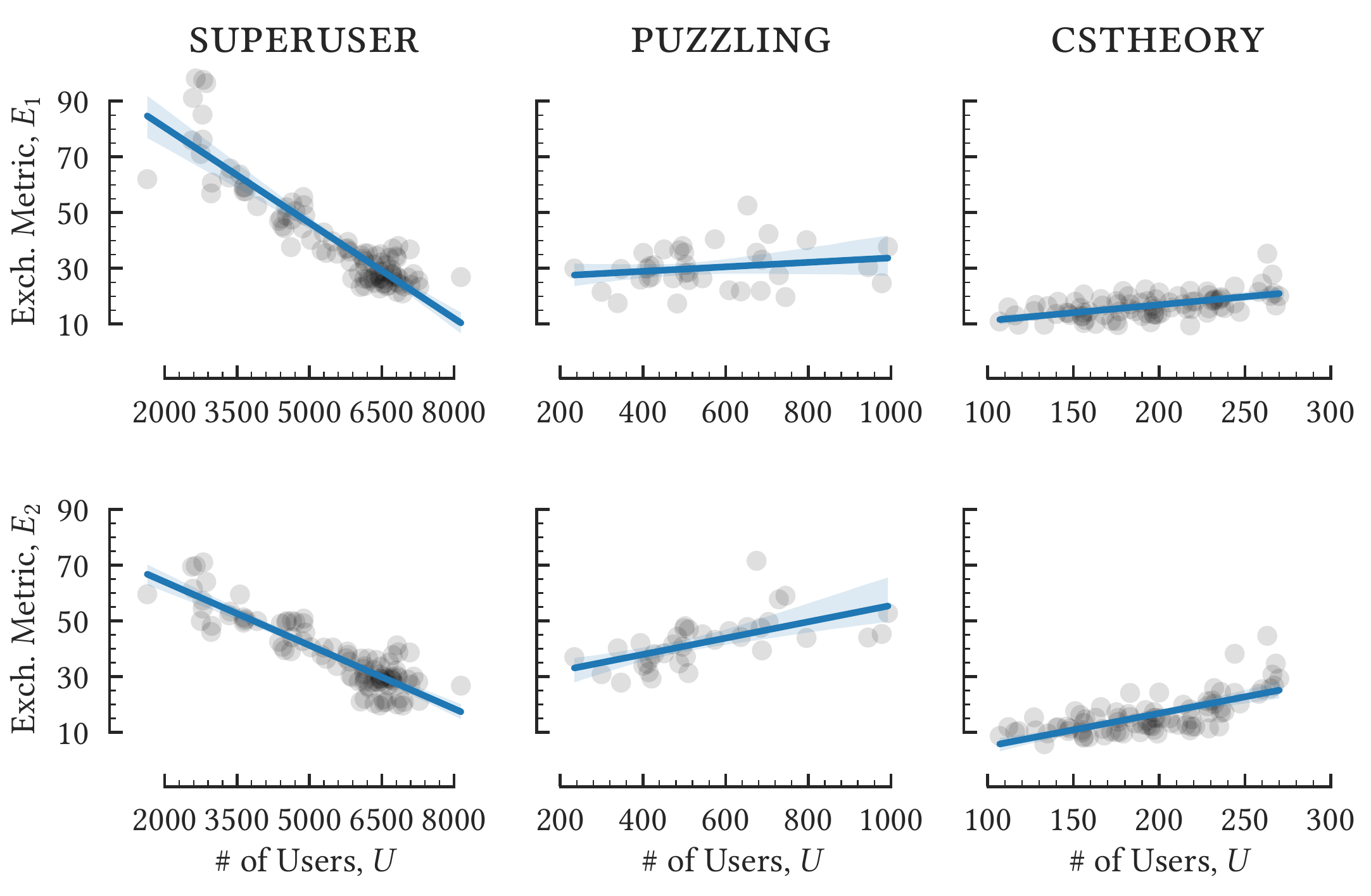}
\caption{User exchangeability under scale: the gap ($E_1$ and $E_2$) between the top contributors and other participants in a knowledge market decreasing or increasing with the increase in number of users. Most markets exhibit a large gap between the top contributors and other participants. Examples: high dissimilarity---\SE{superuser}; moderate dissimilarity---\SE{puzzling}; and low dissimilarity---\SE{cstheory}.}
\label{fig:stability}
\end{figure}

\subsection{Effects on Exchangeability}
Finally, we empirically study the effects of scale on user exchangeability. By exchangeability, we specifically mean the gap between the top contributors and other participants in a knowledge market. Studying this gap is important because it can reveal if a market's success or failure depends on a small group of users.  

To empirically study user exchangeability, we define two metrics that reflect the gap between the top contributors and other participants in a knowledge market. Note that, we only consider the active participants who contributed at least one content. The first metric $E_1$ is defined as the ratio of contribution between the top 5\% and the bottom 5\% of users. For computing $E_1$, we measure the contribution of a user $v$ as the ratio $N^v_{a/q}$ of the number of answers $N^v_{a}$ provided by the user to the number of questions $N^v_{q}$ asked by the user. Notice that $E_1$ is a ratio based metric and we define user contribution to be consistent with this metric. The second metric $E_2$ is defined as the sum of two distances: (i) the distance between the contribution of the top 5\% of users and the median 5\% of users, and (ii) the distance between the contribution of the median 5\% of users and the bottom 5\% of users. For computing $E_2$, we measure the contribution of a user $v$ as a tuple $<N^v_a, N^v_q>$, consisting of the number of answers $N^v_{a}$ provided by the user and the number of questions $N^v_{q}$ asked by the user. Notice that $E_2$ is an interval based metric and we define user contribution to be consistent with this metric. While both metrics have certain limitations, e.g., they are sensitive to outliers, these metrics allow us to comprehend user exchegeability to some extent.

In Figure~\ref{fig:stability} we present the exchangeability of users under scale changes (through $E_1$ and $E_2$) for three \CQA{StackExchange} markets: \SE{cstheory}, \SE{puzzling} and \SE{superuser}. Among the three markets, \SE{superuser} exhibits the highest gap between the top contributors and the other participants. However, as the number of participants increases, this gap decreases, i.e., the users become more exchangeable. In contrast, \SE{cstheory} exhibits the lowest gap between the top contributors and the other participants. However, as the number of participants increase, this gap increases, i.e., the users become less exchangeable.

%% file: 9_discussion.tex
\section{Implications}
Our work promotes two new research directions---size-dependent mechanism design and content dependency in social media---while advancing several others---metrics of market health, power law of participation, and microfoundations of knowledge markets. 

\textbf{Size-Dependent Mechanism Design.} We reveal that the health of a knowledge market depends on the market's size. A natural implication of this dependency is that site operators should adjust mechanisms based on the number of participants. For example, a site operator can decide between retaining existing users (via incentives) and attracting new users (via advertising) based on the number of participants and their activity distribution.

\textbf{Content Dependency in Social Media.} We observe that many social media platforms support several possible user actions with ``complex dependencies''. For example, in Facebook, a post is the root content (primary), comments on the post nest below the post (secondary), and replies to these comments nest beneath the original comments (tertiary). Further, a user can react to any of these content types with several possible reactions. Overall user activity in Facebook is distributed across these possible actions with complex dependencies, which drives the platform's health.

\textbf{Metrics of Market Health.} We demonstrate the presence of diseconomies of scale with several metrics that partially capture the health of a knowledge market. While we concentrate on content-generation based \emph{production metrics}, our concepts can be extended for page-view based \emph{consumption metrics} as well. Also, there is room for developing new health metrics that capture a more detailed picture of a knowledge market's health including \emph{market efficiency}---the degree to which market price (amount of responses and reactions) is an unbiased estimate of the true value of the investment (user effort in content generation)~\cite{damodaran2002}.

\textbf{Power Law of Participation.} In \CQA{StackExchange} markets, a small fraction of the user community participate in high-engagement activities (e.g., linking similar questions), whereas the larger fraction participate in low-threshold activities (e.g., voting). This asymmetry leads to a \emph {Power Law of Participation}~\cite{mayfield2006}. We assert that both low-threshold and high-engagement activities are required for a knowledge market's survival, and should proportionately increase with the increase in number of participants. However, in reality, for most knowledge markets, the size of the user community contributing high-engagement activities does not scale with the system size. This creates a ``gap'' between market supply and demand, and consequently affects market health. 

\textbf{Microfoundations of Knowledge Markets.}
The size-dependent distribution of user contribution implies that users who join a community later in its lifecycle exhibit different behavior than those who were present from the beginning. This very well may imply that the distribution of individual user behaviors (not just their overall production) is ``also'' a function of the system size. We should expect to see a stable user behavior distribution over time for markets that appear to be more scale-insensitive; preliminary results suggest that this may indeed be the case~\cite{geigle2017}.

\section{Limitations}
We discuss several limitations of our work. First, the economic production models do not account for user growth. While there exist several user growth models for two-sided markets~\cite{Kumar2010}, membership based websites~\cite{Ribeiro2014}, and online social networks~\cite{Backstrom2006, kairam2012, zang2016}, it would be useful to introduce economic user growth models that complement our proposed content generation models. Specifically, there is a need to develop resource-based user growth models that account for market health. A potential direction in this research is to extend the Malthusian growth model~\cite{malthus1809}. Second, the proposed production models inherit the fundamental assumptions of macroeconomics: an aggregate is homogeneous (without looking into its internal composition), and aggregates are functionally related etc.~\cite{sims1980}. It would be useful to empirically study these assumptions for real-world knowledge markets.

%% file: x_conclusion.tex
\section{Conclusion}

In this paper, we examined the CQA websites on \CQA{StackExchange} platform through an economic lens by modeling them as knowledge markets. In particular, we designed a set of production models to capture the content generation dynamics in these markets. The resulting best-fit model, Cobb-Douglas, predicts the production of content in StackExchange markets with high accuracy. We showed that the model provides intuitive explanations for content generation. Specifically, it reveals that factors of content generation such as user participation and content dependency have \emph{constant elasticity}; in many markets, factors exhibit \emph{diminishing returns}; markets vary according to their \emph{returns to scale}; and finally many markets exhibit \emph{diseconomies of scale}. We further investigated these prognoses by showing the presence of diseconomies of scale in terms of content production, and several measures of market health. The implications of our work are two-fold: site operators need to design incentives as a function of number of participants; there is a need to develop Economic lenses that can shed insights into the complex dependencies amongst different content types and participant actions in general social networks.